\documentclass[runningheads]{svmult}
\usepackage{makeidx}   
\usepackage{epsfig}
\usepackage{rotate}
\usepackage{subeqnar}  
\usepackage{multicol}  
\usepackage{cropmark} 
\usepackage{physprbb}  
\newcommand{\greeksym}[1]{{\usefont{U}{psy}{m}{n}#1}}
\newcommand{\umu}{\mbox{\greeksym{m}}}

\def\beq{\begin{equation}}
\def\eeq{\end{equation}}
\def\bea{\begin{eqnarray}}
\def\eea{\end{eqnarray}}
\def\bq{\begin{quote}}
\def\eq{\end{quote}}
\def\gappeq{\mathrel{\rlap {\raise.5ex\hbox{$>$}}
{\lower.5ex\hbox{$\sim$}}}}

\def\lappeq{\mathrel{\rlap{\raise.5ex\hbox{$<$}}
{\lower.5ex\hbox{$\sim$}}}}

\def\Toprel#1\over#2{\mathrel{\mathop{#2}\limits^{#1}}}

\begin{document}

\title*{Dark 2002 and Beyond}
\toctitle{Dark 2002 and Beyond}
%
%
\titlerunning{Dark 2002 and Beyond}
%
\author{
\rightline{astro-ph/0204059}
\rightline{CERN-TH/2002-074}
John Ellis
}
\authorrunning{John Ellis}
%
%
\institute{Theoretical Physics Division, CERN\\
CH-1211 Geneva 23}

\maketitle              

\begin{abstract}
Salient aspects of the meeting are summarized, including our knowledge of 
dark matter at different cosmological and astrophysical distance scales,
ranging from large-scale structure to the cores of galaxies, and our 
speculations on particle candidates for dark matter, including neutrinos, 
neutralinos, axinos, gravitinos and cryptons. Comments are also made on 
prospects for detecting dark matter particles and on the dark energy 
problem.
\end{abstract}

The speakers at this conference have come from different backgrounds, from
that of macrophysics - namely astrophysics and cosmology, and from that of
microphysics - namely that of particle physics experiments and theory. We
have come together to discuss a weighty subject Ð namely the nature of
most of the stuff in the Universe. In this brief summary, I shall not be
able to do justice to all the interesting talks we have heard, and I
apologize in advance to those whose work is unjustifiably underemphasized.
Inevitably, my summary adopts personal points of view, that you may not
share.

This talk is ordered according to a sequence of decreasing distance
scales, from the overall size of the Universe at $\sim 10^{10}$~pc down to
the Planck length of $10^{-33}$~cm $\sim 10^{-52}$~pc. Astrophysical
scales that we meet along the way include the $\sim 10^8$~pc of clusters
of galaxies, the $\sim 10^5$~pc of galactic haloes, the $\sim 10^3$~pc of
the cusps of rotation curves, and the $\sim 1$~pc of the central region of
the Milky Way. Among the distance scales of particle-physics experiments
that we meet are the $\sim 10^{-6}$~pc traveled by solar neutrinos, the
$\sim 10^{-10}$~pc travelled by atmospheric neutrinos, the $\sim
10^{-12}$~pc of long-baseline neutrino experiments, and the $\sim
10^{-13}$~pc size of the LHC accelerator. Typical particle sizes we meet
include the Compton wavelength $\sim 10^{-22}$~pc of the neutrino, the
Compton wavelength $\sim 10^{-36}$~pc of supersymmetric particles, and the
characteristic wavelength $\sim 10^{-44}$~pc of the ultra-high-energy
cosmic rays.

Somewhere along this trail, the puzzle of dark matter will surely be
solved, even if we do not yet know where.

\section{Dark Matter on Different Distance Scales}

\subsection{Cosmological Density Parameters}

\begin{figure}[h]
\begin{center}
\includegraphics[width=.9\textwidth]{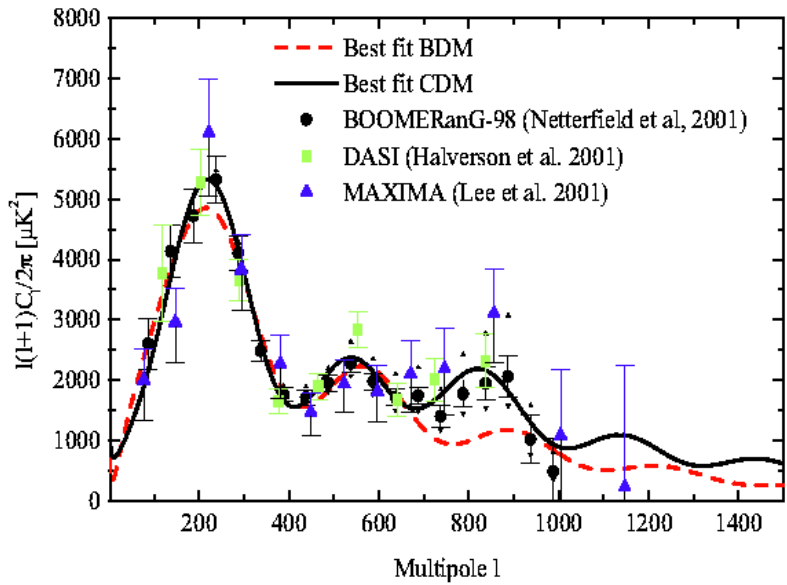}
\end{center}
\caption[]{Recent compilation of data on the cosmic microwave background 
(CMB), exhibiting clearly the first acoustic peak, whose location fixes 
$\Omega_{tot}$, and the possible second and third peaks at larger 
$\ell$~\cite{CMB}.}
\label{fig:CMB}
\end{figure}

\begin{figure}[h] 
\begin{center}
\includegraphics[width=.7\textwidth]{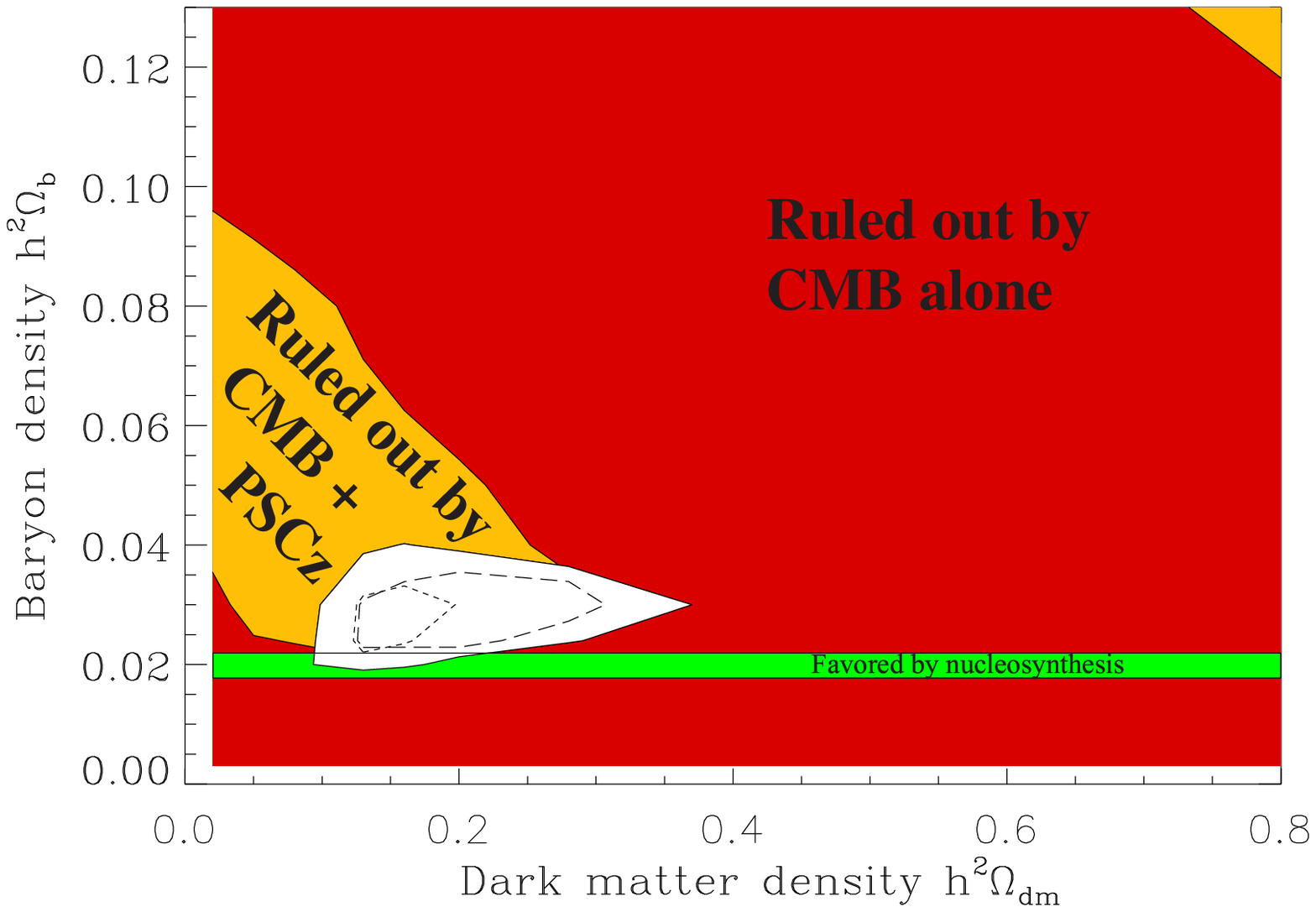}
\end{center}
\caption[]{The combination of cosmological data favours the `concordance' 
model with $\Omega_{tot} \sim 1, \Omega_\Lambda / \Omega_m \sim 2 \to 3$ 
and $\Omega_b$ small, in agreement with cosmological 
nucleosynthesis~\cite{Tegmark}.}
\label{fig:Concordance}
\end{figure}

As we have heard at this meeting, there is abundant evidence for the
dominance of dark matter and dark energy on the largest distance scales.
The cosmological microwave background (CMB) radiation~\cite{CMB} tells us 
that the
total energy density of the Universe, $\Omega_{tot}$, is very close to the
critical value marking the boundary between open and closed 
universes~\cite{Melchiorri}.
This information is provided, in particular, by the value of the multipole
$\ell \sim 210$ at which the first acoustic peak appears in the CMB, 
as seen in Fig.~\ref{fig:CMB}.
This tells us, in effect, the relative sizes of the Universe today and
when the nuclei and free electrons in the primordial plasma combined to
form neutral atoms. There are now indications for a second and even a
third acoustic peak in the CMB at higher $\ell$~\cite{CMB}, but these are 
not yet
securely established. However, the magnitudes of the fluctuations $\delta
T / T$ at these larger values of $\ell$ already tell us that the overall
baryon density $\Omega_b \ll 1$, agreeing to within $ \sim 50$~\% with the
value estimated on the basis of Big Bang nucleosynthesis calculations.
Other information about the large-scale geometry of the Universe for
redshifts $z \lappeq 1$ is provided by data on high-$z$ 
supernovae~\cite{SN}, which
constrain a combination of the matter density $\Omega_m$ and the vacuum
energy density $\Omega_\Lambda$. Combining the CMB and high-$z$ supernova
data, one finds fairly accurate values for the cosmological density
parameters~\cite{Melchiorri}:
\begin{eqnarray}
\Omega_{tot} \; = \; 1.02 \pm 0.06 &,& \; \Omega_m h^2 \; = \;  0.13 \pm
0.05, \; \nonumber \\
\Omega_\Lambda \; = \; 0.5 \pm 0.2 &,& \; \Omega_b h^2 \; = \; 0.022 \pm
0.004,
\label{CMBSNOmega}
\end{eqnarray}
where $h$ is the present-day Hubble expansion rate in units of
100~km/s/Mpc, consistent with the `concordance model': $\Omega_{tot} \sim 
1, \Omega_\Lambda / \Omega_m \sim 2 \to 3$ with $\Omega_b$ 
small, as seen in Fig.~\ref{fig:Concordance}. These data may also be used 
to constrain neutrino 
degeneracy~\cite{Kajino}.
 
There are excellent prospects for significant progress in improving the
CMB constraints using data from the MAP and Planck 
satellites~\cite{Cooray}. One of the
open issues concerns the amount of information likely to be obtained at
large $\ell$~\cite{Aghanim}, particularly from polarization
measurements~\cite{Cooray}. These must contend with weak lensing effects
that must be subtracted in order to extract useful cosmological
information.

\subsection{Large-Scale Structure}

The standard paradigm is that large-scale structures in the Universe are
formed by gravitational instabilities, building on the primordial density
perturbations observed in the CMB~\footnote{Which are commonly thought to
arise from an early epoch of inflation~\cite{Inflation,Hofmann}.}, with 
baryons
falling into the `holes' that are amplified by cold dark matter. Galaxy
formation is considered to be more complex than cluster formation, with
nonlinear astrophysical processes coming into play. Calculating galaxy
formation is therefore more challenging numerically, though it may be
guided by semi-analytical models. The general belief is that clusters
formed before galaxies, which were formed by mergers of smaller
structures.

The observational data on clusters of galaxies support the `concordance
model' in several different ways. Data on X-rays from rich clusters can be
used to estimate the ratio $\Omega_m / \Omega_b$~\cite{Menci}, the results
suggesting again that $\Omega_m \ll 1$. The evolution of large-scale
structure as a function of redshift also supports the concordance model,
there being many more clusters at high $z$ than would be expected in an
Einstein-de-Sitter cosmology with $\Omega_m \sim \Omega_{tot} \sim
1$~\cite{Guzzo}.  Moreover, the shape of the two-point correlation
function for galaxy clusters, as measured by the REFLEX collaboration,
agrees with the concordance model, and is very similar in shape to that of
the two-point correlation function for galaxies. The overall normalization
of the cluster correlation function is considerably higher, as expected
from the `biasing' phenomenon, according to which rarer peaks are
correlated more strongly than those of lower significance~\cite{Guzzo}.

Several large new surveys of galaxy redshifts are underway, with some
initial results from samples of $\sim 10^5$ galaxies now becoming
available. Results from the first release of Sloane Digital Sky Survey
(SDSS) data (29,000 redshifts) agree very well with those from the 2dF
(200,000 redshifts) team and from the LCRS data (26,000 redshifts). Once
again, the shape of the two-point correlation function agrees very well
with the concordance model, and disagrees with the Einstein-de-Sitter
model.

\begin{figure}[h]
\begin{center}
\includegraphics[width=.6\textwidth]{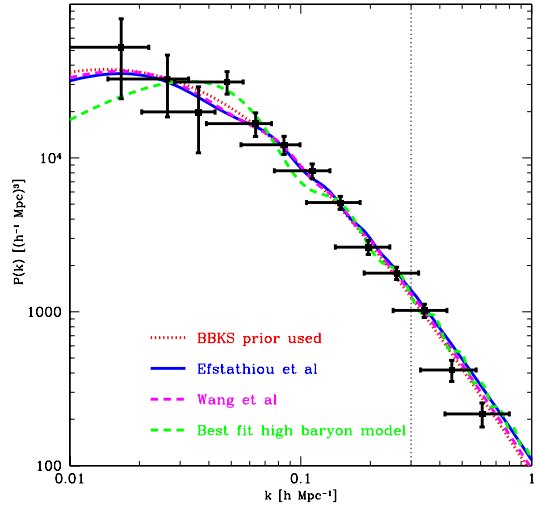}
\end{center}
\caption[]{A recent reanalysis~\cite{Hamilton} of the data 
of~\cite{Peacock} does not reveal any baryonic `wiggles'.}
\label{fig:wiggles}
\end{figure}

Just as the CMB is expected to display several acoustic peaks, so also the
galaxy correlation function is expected to exhibit baryonic `wiggles' on
scales up to $\sim 100$~Mpc. There was recently a claim~\cite{Peacock} to
have observed an indication of them, but the amplitude was unexpectedly
large, and a more recent reanalysis casts doubt on the `wiggle'
interpretation~\cite{Hamilton}, as seen in Fig.~\ref{fig:wiggles}.

Models of structure formation are also constrained by data on quasars and
Lyman $\alpha$ forests. The numbers of high-$z$ quasars are problematic
for hot- and warm-dark-matter models, but are consistent with the
concordance model. The data on Lyman $\alpha$ forests are particularly
interesting, because reduced nonlinearities render their interpretation
less ambiguous than those of galaxies.

As repeatedly mentioned, all these data are consistent with the 
concordance
model. Combining them with the CMB and high-$z$ supernova data, one
obtains the refined estimates:
\begin{eqnarray}
\Omega_{tot} \; = \; 0.99 \pm 0.03 &,& \; \Omega_m h^2 \; = \;  0.14 \pm
0.02, \; \nonumber \\
\Omega_\Lambda \; = \; 0.65 \pm 0.05 &,& \; \Omega_b h^2 \; = \; 0.021 \pm
0.003.
\label{AllOmega}
\end{eqnarray}
However, there are some open issues in structure formation. One is that
walls of galaxies are seen in the data that do not appear in simulations,
although this is not seen as a serious problem. Another potential problem 
is
provided by density profiles in groups and clusters of galaxies, but here
recent simulations now indicate that energy ejection due to supernovae
breaks self-similarity in a manner consistent with observation, within the
concordance model. There have also been worries about the numbers of
satellite galaxies, with many fewer being observed than are expected in
the concordance model. However, there are reasons to think that small
galaxies might not light up~\cite{Moore}, since their conventional matter 
could be
dispersed by the radiation from the first generation of stars.

\subsection{Dynamics of Cusps}

Another potential problem is whether cold dark matter predictions of cusps
in the cores of galaxies are compatible with observations~\cite{Menci}. 
The cold dark
matter density in a generic spheroidal galactic halo may be parameterized
by
\begin{equation}
\rho (r) \; = \; {r_0^\gamma \over r^\gamma (1 + ({r \over
r_0})^\alpha)^{(\beta - \gamma) \over \alpha}},
\label{rhohalo}
\end{equation}
where $r_0$ is some scale factor. Early isothermal models of galactic
haloes were nonsingular, with $\gamma = 0, \alpha = \beta = 2$. These were
superseded by Navarro-Frenk-White profiles~\cite{NFW} with central 
singularities:
$\gamma = 1, \alpha = 1, \beta = 3$, and more recently by even more
singular profiles: $\gamma = 1.5, \alpha = 1.5, \beta = 3$~\cite{Merritt}. 
Observations
do not support such singular cusps, and various attempts have been made to
understand whether and how they might be weakened.

One suggestion has been that the annihilations of dark-matter particles in
the cusps might generate particles and radiation pressure that would
disperse the cusps. However, it seems questionable whether the
annihilation rates found in plausible models are large enough, and the
annihilations are so constrained by upper limits on synchrotron radiation
that this mechanism seems unlikely to work~\cite{Merritt}.

\begin{figure}[h]
\begin{center}
\includegraphics[width=.6\textwidth]{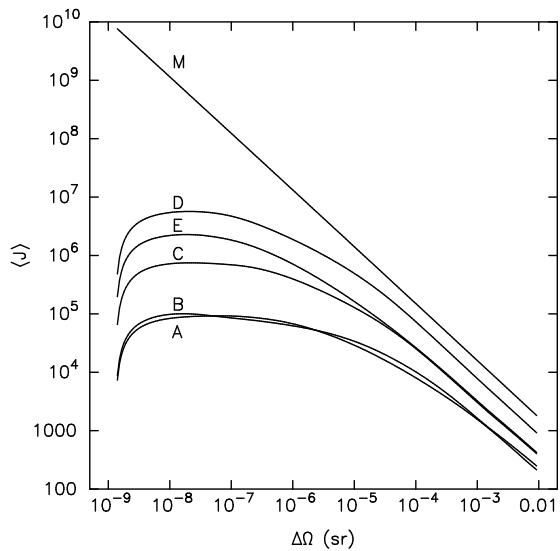}
\end{center}
\caption[]{The gravitational slingshot effect may suppress the density
in a galactice core: the curves illustrate the reductions 
in the photon flux from relic annihilation, integrated over the 
indicated angular range, as found in different scenarios~\cite{Merritt}.}
\label{fig:Merritt}
\end{figure}

Another suggestion made at this meeting, that seems more promising, is
that black holes at the centers of galaxies~\cite{Madejski} may disrupt 
the cusps via the
gravitational slingshot effect acting on individual dark-matter 
particles~\cite{Merritt}, as seen in Fig.~\ref{fig:Merritt}.
This effect could be important during mergers, and simulations indicate
that a further suppression of the core density would occur if there is a
hierarchy of mergers. It does not seem at the moment that the apparent
absence of cusps in generic galaxies is necessarily a big problem for cold
dark matter.

A more specialized question concerns the center of the Milky Way. As
discussed in more detail below, this is known to contain a central massive
object weighing $\sim 3 \times 10^6$~solar masses, that is normally
presumed to be a black hole~\cite{Eckart}. Like any other galaxy, one
would expect the history of our own to have included a number of mergers,
that are likely to have suppressed the central density spike. For this
reason, the density of cold dark matter at the center of the Milky Way,
and hence the rate of their annihilations, has a considerable uncertainty
that must be borne in mind when assessing dark-matter detection
strategies, as discussed later.

\subsection{Matter Content of the Milky Way}

As we have been reminded at this meeting, one form of dark matter has been
discovered, namely MACHOs~\cite{Griest}. It is still unclear what fraction
of the microlensing events is due to MACHOs in our galactic halo, and what
fractions are due to objects in either the Magellanic Clouds or our own
(extended) galactic disc. If all the observed microlensing events
originate from our galactic halo, as much as 8~\% to 40~\% of it could be
composed of MACHOs. However, it now seems clear that MACHOs cannot
constitute the bulk of the halo of the Milky Way~\cite{Drake}.

\begin{figure}[h]
\begin{center}   
\includegraphics[width=.7\textwidth]{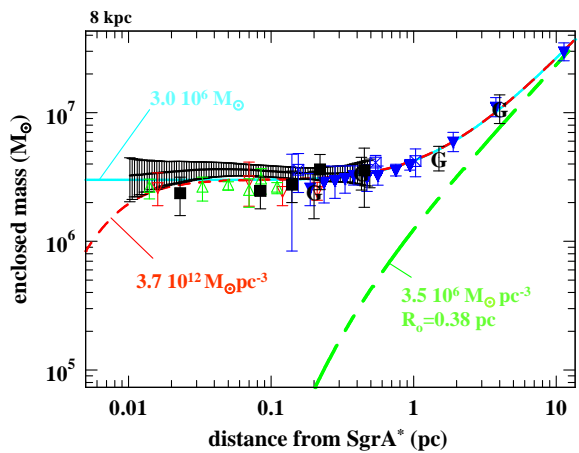}
\end{center}
\caption[]{The centre of the Milky Way contains a heavy 
object Sagittarius A*, with a mass $\sim 3 \times 10^6$~solar 
masses concentrated in a small radius~\cite{Eckart}.}
\label{fig:Hole}
\end{figure}

\begin{figure}[h]
\begin{center}
\includegraphics[width=.4\textwidth,angle=270]{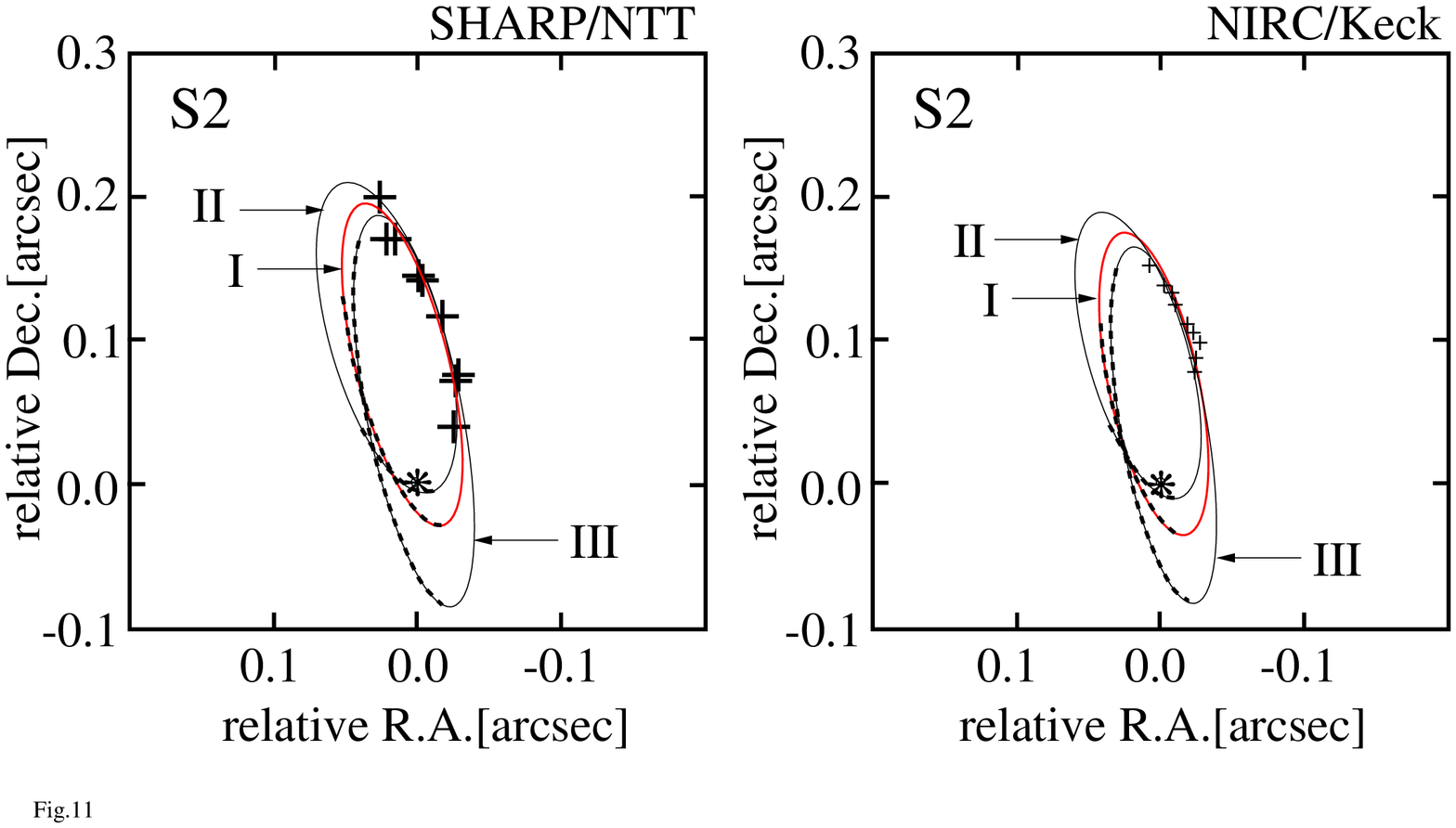}
\end{center}
\caption[]{Orbit of the star S2 near the heavy object Sagittarius 
A* at the centre of the Milky Way, exhibiting clearly the 
curvature due to its gravitational attraction~\cite{Eckart}.}
\label{fig:Eckart}
\end{figure}

A topic discussed at length at this meeting has been the composition of
the central object in our galaxy, Sagittarius A*, that weighs $\sim 3
\times 10^6$~solar masses~\cite{Eckart}, as seen in Fig.~\ref{fig:Hole}. 
Although this is normally
presumed to be a black hole, but this has not been established. We know
from the observation of adjacent stellar orbits that the central mass must
be concentrated within a small radius. Curvature has been
observed in the orbits of some nearby stars, as seen in 
Fig.~\ref{fig:Eckart}, and the rate of precession of
these orbits promises to become a useful tool for measuring how much of
the mass of Sagittarius A* is extended.

A point in favour of the black hole interpretation of Sagittarius A* is
that it has been observed to flare in X-rays, exhibiting large luminosity
variations over very short time scales~\cite{Baganoff}. Matter falling
into a black hole is expected to emit X-rays, and the rapid time variation
indicates that the central engine of Sagittarius A* must be small.
However, Sagittarius A* is not very bright, and it has been argued that
this poses a so-called `blackness problem' which motivates considering
other models. However, as was discussed here~\cite{Eckart}, the blackness 
of Sagittarius
A* is not necessarily a problem, since there are advective flow models
that appear well able to reproduce its observed brightness~\cite{Narayan}.

\begin{figure}[h]
\begin{center}
\includegraphics[width=.4\textwidth]{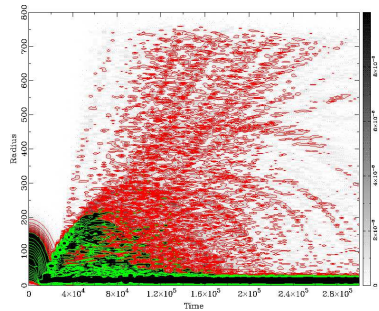}
\end{center}
\caption[]{Illustration of the formation of a fermion 
ball~\cite{Lindebaum}, showing the initial infall and the subsequent 
bouncing of material, some of which escapes while most falls back.}
\label{fig:Lindebaum}
\end{figure}

As alternatives to the black hole hypothesis, balls of condensed bosons or
fermions~\cite{Lindebaum,Chavanis,Bilic,Viollier} have been proposed as
alternative models for Sagittarius A*. The latter model postulates a
neutral, weakly-interacting fermion weighing about 15~KeV, and we have
seen here detailed simulations of the evolution with time of a ball made
out of such fermions~\cite{Lindebaum}, as seen in 
Fig.~\ref{fig:Lindebaum}. This could not be a conventional
neutrino, because the oscillation experiments tell us that they are
degenerate to within $10^{-2}$~eV$^2$, and Tritium $\beta$-decay
experiments tell us that the $\nu_e$ mass is less than about 2.5~eV.
Moreover, astrophysical and cosmological data suggest a similar upper
limit on all the neutrino species. We also heard how such a fermion might
also constitute the halo of the Milky Way~\cite{Bilic}.

A potential `smoking gun' for the black-hole interpretation of Sagittarius
A* is a Fe emission line at about 8.6~KeV, that is expected to be smeared
by the gravitational redshift near the horizon. An effect consistent with
this expectation has been reported. On the other hand, radiative decay of
a fermion weighing 15 to 17~KeV could also produce photons in the same
energy range!

In my view, there is no reason to abandon the conservative black-hole
paradigm for Sagittarius A*, and Occam's razor prompts me to favour it. On
the other hand, such a paradigm must be challenged constantly, and it is
good to have a rival model that we can use to benchmark the success of the
black-hole paradigm.

\section{Particle Candidates for Dark Matter}

Now that the astrophysicists have convinced us of the necessity of dark
matter, what candidates have we particle physicists to offer? {\it
Neutrinos} are the prime candidates for hot dark matter, whereas there are
numerous candidates for cold dark matter, including {\it axions},
supersymmetric particles such as the lightest {\it neutralino}, the {\it
axino} and the {\it gravitino}, and possible superheavy metastable relics
such as {\it cryptons}. Axions were not much discussed here, so I
concentrate on neutrinos, supersymmetric particles and superheavy relics.

\subsection{Neutrinos}

As we were reminded at this meeting, alternative interpretations of the
solar and atmospheric neutrino data are not excluded, but oscillations
between different neutrino mass eigenstates are very much the favoured
interpretations~\cite{Valle}.

In the case of atmospheric neutrinos, the favoured oscillation pattern is
$\nu_\mu \to \nu_\tau$. There is no evidence for oscillations involving
the $\nu_e$, there being no anomaly in the atmospheric $\nu_e$ data, and
we also have a stronger upper limit on $\nu_e \to \nu_\mu$ oscillations
from the Chooz data. Also, $\nu_\mu \to \nu_{sterile}$ oscillations are
disfavoured by the zenith-angle distributions. The central value of the 
mass-squared difference $\Delta
m^2 \sim 2.5 \times 10^{-3}$~eV$^2$, and the mixing angle is large:
$\sin^2 2 \theta > 0.8$~\cite{superK}.

\begin{figure}[h]
\begin{center}
\vspace*{1.5in}
\hspace*{1.25in}
\includegraphics[width=.6\textwidth]{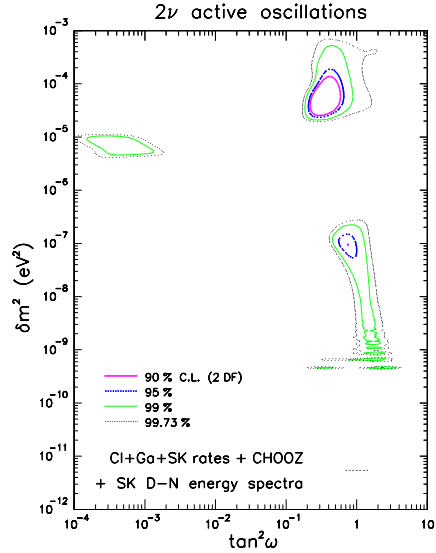}
\end{center}
\caption[]{Regions of 2-neutrino oscillation parameters 
allowed~\cite{Lisi} by the available
solar neutrino data, including those from SNO~\cite{SNO}.}
\label{fig:Lisi}
\end{figure}

In the case of solar neutrinos, $\nu_e \to \nu_\mu$ and/or $\nu_\tau$
oscillations are preferred, though some admixture of $\nu_e \to
\nu_{sterile}$ cannot be excluded. After SNO~\cite{SNO}, the data 
increasingly
favour the large-mixing-angle (LMA) solution, with $10^{-5}~{\rm eV}^2 <
\Delta m^2 < 10^{-4}~{\rm eV}^2$, though large mixing with a somewhat
smaller value of  $\Delta m^2$ (the LOW solution) is also 
possible~\cite{Lisi}, as seen in Fig.~\ref{fig:Lisi}.

We heard at this meeting of a possible indication for neutrinoless
double-$\beta$ decay, corresponding to $0.11~{\rm eV} < \; 
<m>_{\beta \beta} \; <
0.56~{\rm eV}$~\cite{KKG}~\footnote{The possible interpretation in the
context of oscillation experiments was also discussed
here~\cite{Minakata}.}.  The claimed significance is not high, around 2
$\sigma$, whereas 4 or 5 $\sigma$ would be needed to claim a discovery.
The claimed indication rests on the interpretation of one possible bump in
the energy spectrum, with some other bumps being interpreted as other
radioactive decays, and the rest resisting interpretation. However, the
analysis has been criticized~\cite{anti}, in particular because the 
strengths of the
other claimed radioactive decays appear to be too high. The
Heidelberg-Moscow experiment has now reached the limit of its sensitivity,
so we must wait to see what other experiments find. In a conventional
hierarchical scenario for neutrino masses, one could expect to see a
signal at the level $<m>_{\beta \beta} \sim 0.01$~eV, for which a future
large-scale experiment such as the GENIUS proposal will be needed.

\begin{figure}[h]
\begin{center}
\includegraphics[width=.7\textwidth]{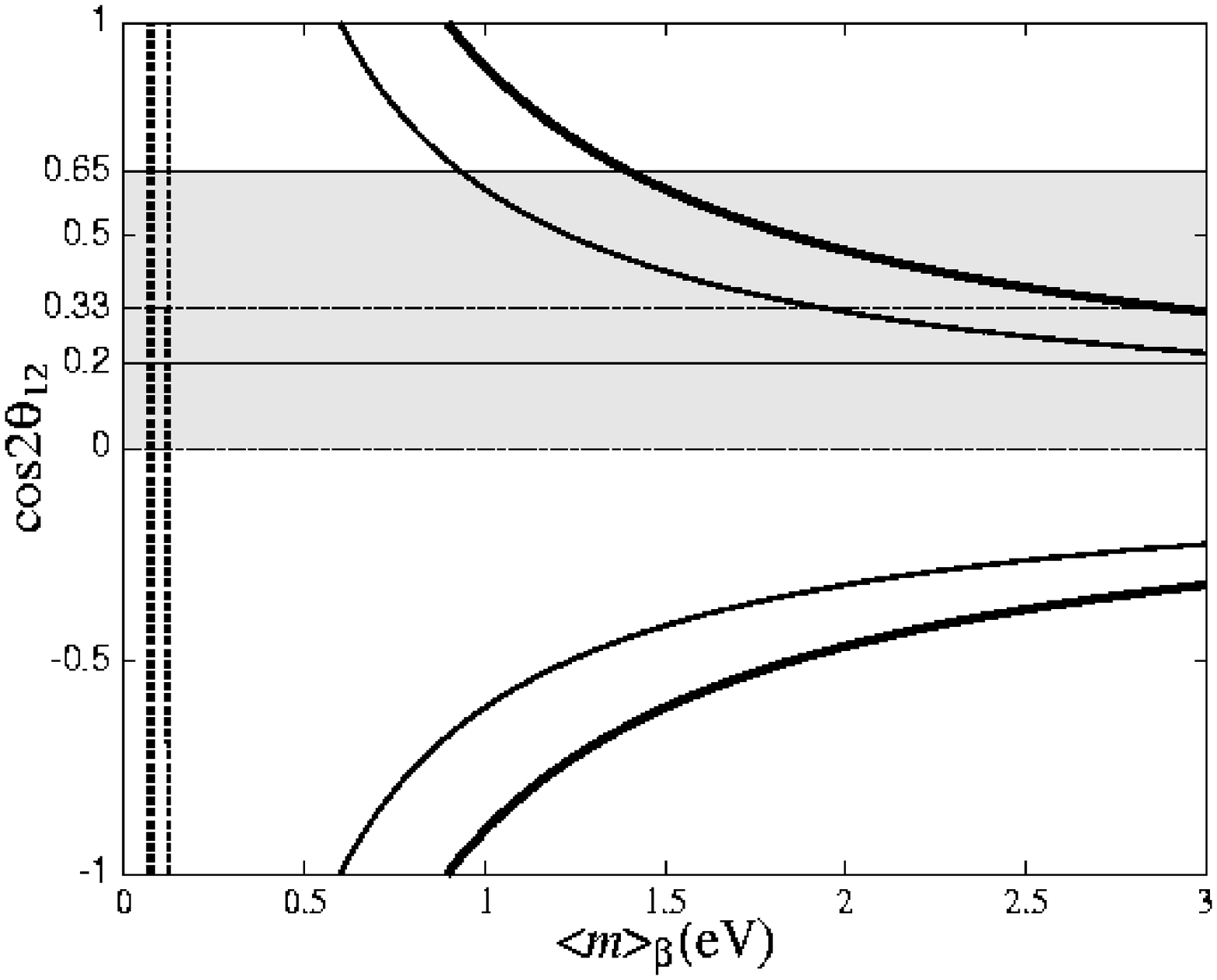}
\end{center}
\caption[]{The constraint on the neutrino mixing angle $\theta_{12}$ and 
the Tritium $\beta$-decay observable $<m>_\beta$ that would be imposed if
$0.11~{\rm eV} < \; <m>_{\beta \beta} \; < 0.56~{\rm eV}$ (thin lines) 
or $0.05~{\rm eV} < \; <m>_{\beta \beta} \; <
0.84~{\rm eV}$ (thick lines)~\cite{Minakata}. The ranges of $\theta_{12}$ 
favoured by the LMA (LOW) solutions are shaded between thin solid (dashed) 
lines.}
\label{fig:Minakata}
\end{figure}

Tritium $\beta$-decay experiments are largely complementary to
neutrinoless double-$\beta$ decay experiments~\cite{Otten}, since they 
measure a different observable:
\begin{equation}
<m>_\beta \; \equiv \; \Sigma_i | U_{ei}|^2m_i \; {\rm vs} \;
<m>_{\beta \beta} \; \equiv \; | \Sigma_I U_{ei}^2 e^{i \phi_i} m_i |.
\label{different}
\end{equation}
As we heard at this meeting, the previous problems of these experiments,
namely the tendency to prefer negative values of  $<m>_\beta^2$ and the
appearance of a `bump' near the end of the spectrum, have now been
resolved. The former has been traced to a roughening transition in the
frozen Tritium surface layer, and the latter to plasma effects related to
particles trapped in the spectrometer. The limits accessible with the
present experiments have now been almost saturated. Each experiment
reports an upper limit $\sim 2.2$~eV, but a more conservative
interpretation would be
\begin{equation}
<m>_\beta < 2.5~{\rm eV}.
\label{limitmeff}
\end{equation}
As we also heard at this meeting, there are ambitious ideas for a
next-generation experiment called KATRIN~\cite{Steidl}, which is proposed 
to be 70~m
long and 8~m in diameter, and able to reach a limit of 0.35~eV. This is
certainly a worthwhile objective, but the question still remains: how to
reach the atmospheric neutrino mass scale $\sim 0.05$~eV.

Combining the upper limit (\ref{limitmeff}) from Tritium $\beta$ decay
with the upper limit $<m>_{\beta \beta} < 0.56$~eV from $\beta 
\beta_{0\nu}$ decay, and
recalling the lower limits: $\Delta m^2 > 0.003$~eV$^2$ from atmospheric
neutrinos, $10^{-4}$~eV$^2$ for the LMA solar solution, one can infer the
following allowed range for the total relic neutrino density:
\begin{equation}
~0.001 < \; \Sigma_\nu \Omega_\nu \; < \; 0.18.
\label{Omeganu}
\end{equation}
Though not conclusive, this is certainly consistent with the lack of
enthusiasm for hot dark matter indicated by studies of large-scale
structure. The CMB is relatively
insensitive to $\Omega_\nu$, whereas large-scale structure is very
sensitive to $\Omega_\nu$~\cite{Tegmark}.

We can look forward to significant advances in neutrino physics in the
coming years that will check the emergent picture outlined above. The SNO
experiment will soon provide important new data on the ratio of neutral-
and charged-current events~\cite{SNO}. Also starting to take data is the
KamLAND experiment~\cite{KamLAND}, which should provide a conclusive test
of the LMA interpretation of the solar neutrino data. Shortly, we can also
expect important data from BOREXINO~\cite{BOREXINO}, that will also help
pin down the interpretation of solar neutrinos.

Long-baseline neutrino experiments~\cite{Zuber} are now swinging into
action to probe the interpretation of atmospheric neutrinos, led by 
K2K~\cite{K2K}.
The Super- KAMIOKANDE detector is currently being reconfigured after its
accident, so as to enable data-taking for the K2K experiment to restart.
Meanwhile, the NUMI beam and the MINOS detector are being prepared in the
United States~\cite{MINOS}, in parallel with the CNGS beam and the OPERA 
detector in
Europe~\cite{OPERA}. The MINOS experiment is aimed at measurements of the 
oscillation
pattern, neutral- and charged-current rates, and the search for $\nu_e$
appearance in a $\nu_\mu$ beam, whilst the OPERA detector is aimed at
detecting $\tau$ production in a $\nu_\mu$ beam due to $\nu_\mu \to
\nu_\tau$ oscillations. In the longer term, the JHF is under 
construction~\cite{JHF},
and will provide the opportunity to generate a more intense neutrino beam
that could be directed towards the SuperKAMIOKANDE detector or its
projected megaton-class successor, HyperKAMIOKANDE. The latter would
provide a first opportunity to search for CP neutrino oscillations.

The search for CP violation could be made much more precise if more
intense, pure $\nu_e$ and/or $\nu_\mu$ were available. A relatively new
idea to realize this objective is to store radioactive ions whose decays
would yield pure $\nu_e$ and ${\bar \nu}_e$ beams~\cite{Zucchelli}. A
longer-standing concept is that of a neutrino factory based on the decays
of muons in a storage ring~\cite{nufact}. Since this produces
simultaneously $\nu_\mu$ and ${\bar \nu}_e$ beams, both with
well-understood spectra, a neutrino factory is the ultimate weaqpon for
neutrino-oscillation studies. Among the objectives of this programme would
be the ultimate searches for $\theta_{13}$ and CP violation~\cite{CPnu},
and determining the sign of the neutrino mass hierarchy.

In addition to neutrino-oscillation studies, a neutrino factory would also
offer interesting prospects for high-statistics studies of neutrino
interactions with a short-baseline beam~\cite{SBL}. The intense proton 
source would
provide other particle physics opportunities, for example using stopped or
slow muons~\cite{slowmu}, as well as opportunities in other areas of 
physics~\cite{Kfact}. Many
accelerator laboratories around the world, including Europe, Japan and the
United States, have initiated studies of neutrino factories. However, in
each region it seems that the first choice for a major new accelerator
facility is a linear $e^+ e^-$ collider. Thus there is a danger that the
neutrino factory will be `always the bridesmaid, never the bride'. The
priority accorded a linear $e^+ e^-$ collider is understandable, but a
balanced accelerator programme for the world should surely also include a
neutrino factory somewhere. Both machines could cast important light on 
the dark matter problem, in different ways.

\subsection{Supersymmetric Dark Matter}

Assuming that $R \equiv (-1)^{B+L+2S}$ is conserved, the lightest
supersymmetric particle (LSP) is stable, and may be an ideal candidate for
cold dark matter, provided it is neutral and has no strong interactions. 
The possibility most often studied~\cite{EHNOS} is that 
the LSP is
the lightest {\it neutralino} $\chi$, a mixture of the supersymmetric
partners of the photon, $Z^0$ boson and neutral Higgs bosons. Another
option mentioned here is that the LSP is the {\it axino} $\tilde a$, the
supersymmetric partner of the hypothetical axion. Finally, there is the
{\it gravitino} $\tilde G$, which is generally unwelcome, since detecting
it would be very difficult. The gravitino option was not discussed here,
so I do not discuss it either. Instead, I focus mainly on the lightest
neutralino $\chi$, mentioning more briefly the axino option.

\subsubsection{Neutralino Dark Matter}

\begin{figure}[h]
\begin{center}
\includegraphics[width=.7\textwidth]{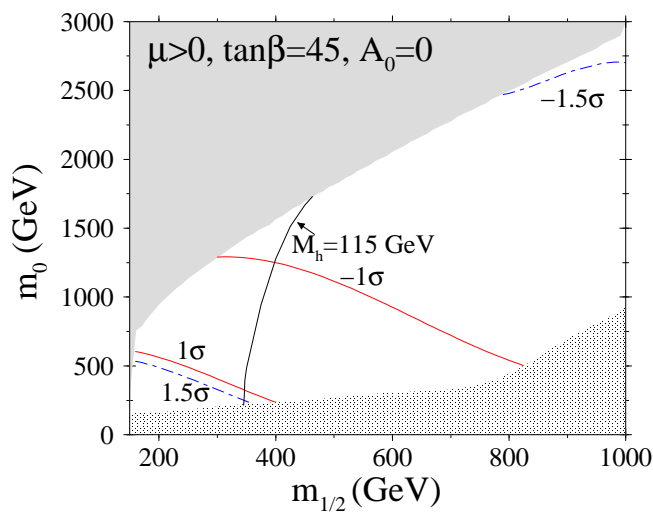}
\end{center}
\caption[]{Compilation of limits on the constrained MSSM for $\mu > 
0, \tan \beta = 45, A_0 = 0$, showing in particular 
the region favoured by $g_\mu - 2$~\cite{Nath} at the $\pm 1, 
1.5$-$\sigma$ levels. Electroweak symmetry breaking is not posible in the 
shaded region at the top, and the lightest supersymmetric particle would 
be the lighter $\tilde \tau$ in the hatched region at the bottom.}
\label{fig:Nath}
\end{figure}

\begin{figure}[h]
\begin{center}
\vspace*{0.5in}
\includegraphics[width=.4\textwidth]{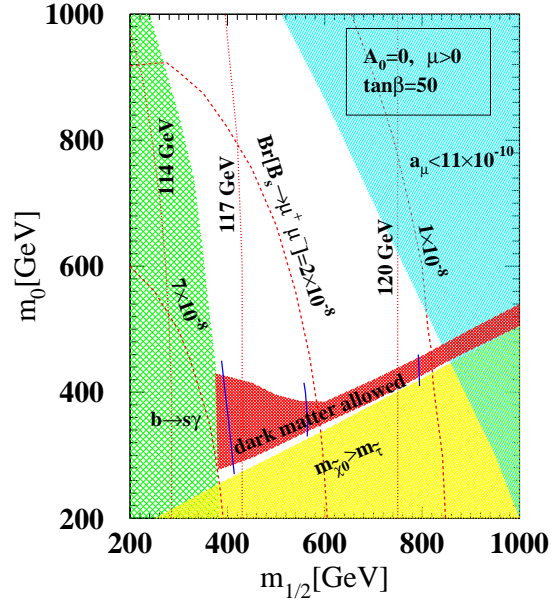}
\end{center}
\vspace*{1in}
\caption[]{Compilation of limits on the constrained MSSM for $\mu >
0, \tan \beta = 55, A_0 = 0$, showing in 
particular the region accessible to the search for $B_s \to \mu^+ \mu^-$ 
at the Tevatron collider~\cite{Arnowitt}. Regions on the left and 
right 
are disfavoured by $b \to s \gamma$ and $g_\mu - 2$, respectively. In 
the hatched region at the bottom, the lightest supersymmetric particle 
would be the lighter $\tilde \tau$. The allowed region for 
supersymmetric dark matter is shaded.} 
\label{fig:Arnowitt}
\end{figure}

In some sense, the neutralino is the most `natural' candidate in the
minimal supersymmetric extension of the Standard Model (MSSM), since one
normally expects it to be lighter than the the gravitino in models based
on supergravity, and a relic density in the range of interest to
astrophysicists and cosmologists: $0.1 < \Omega_\chi h^2 < 0.3$ is
`generic'. As several speakers have shown here, neutralino dark matter is
compatible with all the available accelerator constraints, including
searches for supersymmetric particles at LEP~\cite{Rosier-Lees},
HERA~\cite{meyer} and the Tevatron collider, as well as the indirect
constraints imposed by measurements of $b \to s \gamma$ and $g_\mu -
2$~\cite{Spanos,Ellis,Nath,Arnowitt,Roszkowski}. In the most constrained
versions of the MSSM, in which scalar and fermionic sparticle masses are
each universal at some input grand-unification scale, as in simple
supergravity (SUGRA) models, the lightest neutralino probably weighs more
than about 100~GeV~\cite{EFGOSi}. Figs.~\ref{fig:Nath} and 
\ref{fig:Arnowitt} show examples of the allowed parameter space 
in the constrained MSSM, illustrating the range allowed by the $g_\mu -
2$ constraint at the 1.5-$\sigma$ level, and the potential power of the 
search for $B_s \to \mu^+ \mu^-$ at the Tevatron collider, respectively.

\begin{figure}[h]
\begin{center}
\includegraphics[width=.7\textwidth]{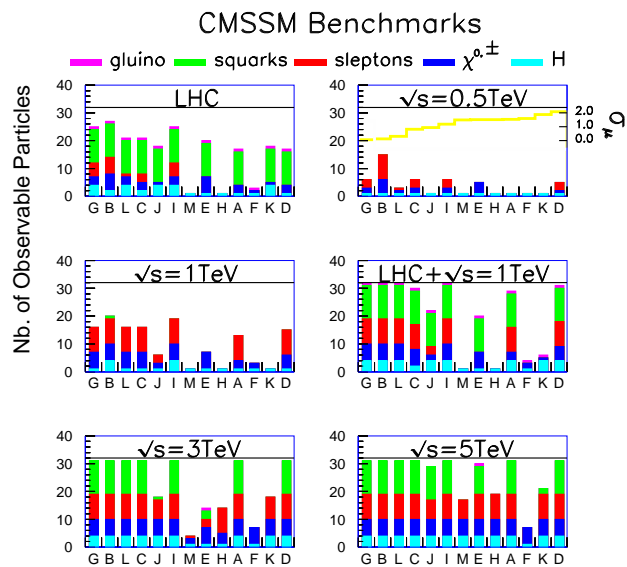}
\end{center}
\caption[]{
Summary of the prospective sensitivities of the LHC and
lepton colliders with
different $\sqrt{s}$ energies to CMSSM particle production
in the proposed benchmark scenarios~\cite{Bench}, which are
ordered by their distance from the central value of $g_\mu - 2$, as
indicated by the pale (yellow) line in the second panel. We see clearly 
the complementarity between a lepton ollider and the LHC in the TeV
range of energies~\cite{Bench}, with the former
excelling for non-strongly-interacting particles, and the LHC for
strongly-interacting sparticles.}
\label{fig:Manhattan}
\end{figure}

The allowed parameter space may be explored theoretically either by
parameter scans, or by focusing on specific benchmark scenarios intended
to illustrate the range of possibilities left open by the experimental
constraints~\cite{Bench}. These indicate that dark matter searches can
expect strong competition from future accelerators, notably the LHC. This
will be able to explore much of the domain of parameters allowed by the
relic density constraint and current experimental constraints. Moreover,
as has been revealed by specific benchmark studies illustrtaed in
Fig.~\ref{fig:Manhattan}, in much of the accessible parameter space the
LHC may be able to discover several different types of supersymmetric
particles, and measure the CMSSM parameters quite accurately. 
However, as
also shown in Fig.~\ref{fig:Manhattan}, there are some benchmark scenarios
where the LHC does little more than discover the lightest MSSM Higgs
boson. Experiments searching for dark matter have an almost clear field
until 2007, but will then get some serious competition: caveat the LHC!

\subsubsection{Neutralino Relic Density Calculations}

These often assume universal input scalar masses, termed here the CMSSM,
as found in minimal supergravity (mSUGRA) models. In the CMSSM, there is a
`bulk' region of relatively low values of $m_{1/2}, m_0$ where the relic
density falls with in the range $0.1 < \Omega_\chi h^2 < 0.3$ favoured by
astrophysics and cosmology. Stretching out from the bulk region to larger
$m_{1/2}$ and/or $m_0$ are filaments of parameter space where special
circumstances suppress the relic density, in some of which the LSP mass
$m_\chi$ may be significantly heavier. These filaments may appear because
of coannihilation~\cite{coann} - in which the relic LSP density is
suppressed by mutual annihilations with other sparticles that happen to be
only slightly heavier, rapid annihilation through direct-channel boson
resonances - in particular the heavier neutral MSSM Higgs
bosons $A, H$, and in the `focus-point' region~\cite{focus} near the
boundary where calculations of electroweak symmetry breaking fail.

The relic density in the bulk region is relatively insensitive to the
exact values of the input CMSSM parameters, and to differences in the
(inevitable) approximations made in the calculations~\cite{EO}. However,
relic-density calculations in the filament regions are much more sensitive
to these input values and approximations, and hence more likely to differ
from one paper to another, as we have seen at this meeting. Several
different codes for calculating the relic density are now available, and
the most recent ones generally agree quite well, once the differences in
inputs and approximations are straightened out.

\subsubsection{Strategies to Search for Neutralinos}

\begin{figure}[h]
\begin{center}
\includegraphics[width=.5\textwidth]{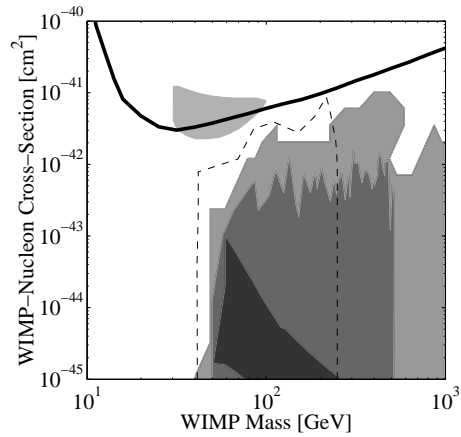}
\end{center}
\caption[]{The elastic scattering cross section possible 
in the constrained MSSM~\cite{Spanos,Roszkowski,EFlO} lies considerably 
below the 
range suggested by DAMA~\cite{Belli}, which is not yet excluded by 
CDMS~\cite{CDMS}.}
\label{fig:CDMS}
\end{figure}

The most direct signal for supersymmetric dark matter would be {\it
scattering on nuclei}~\cite{GW}, a topic discussed by many speakers at
this meeting.  The observation of an annual modulation effect in the DAMA
detector was reported here~\cite{Belli}, but the source of the modulation
has not yet been pinned down. Detectors using other techniques have not
yet been able to confirm the DAMA results~\cite{CDMS}, but neither have 
they yet been
ruled out. Most calculations now agree that it is very difficult to
reproduce in the constrained MSSM the elastic scattering cross section 
that would be
required by DAMA~\cite{Spanos,Roszkowski,EFlO}, as seen for example in 
Fig.~\ref{fig:CDMS}.

\begin{figure}[h]
\begin{center}
\hspace*{-0.2in}
\includegraphics[width=.5\textwidth]{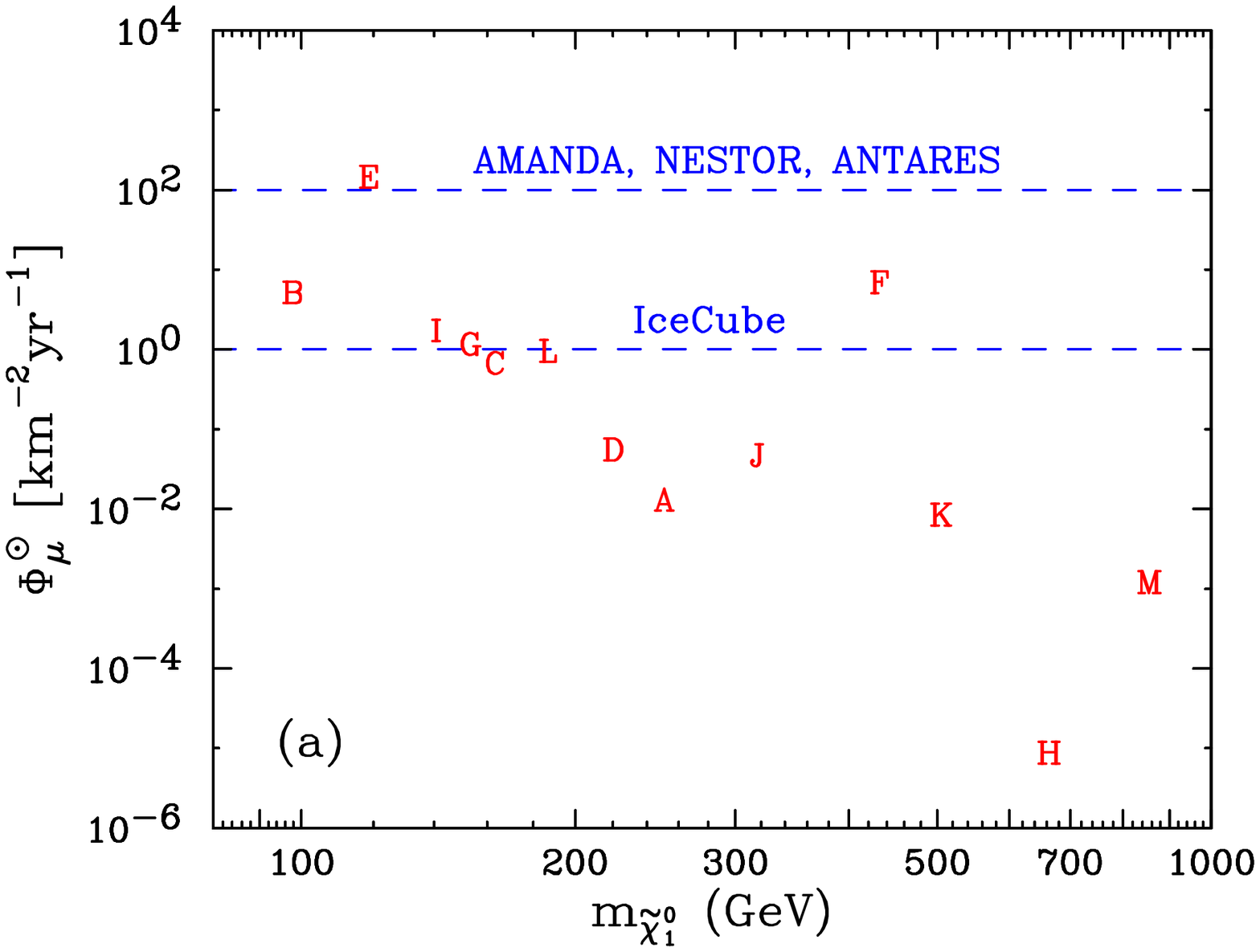}
\hspace*{-0.1in}
\includegraphics[width=.5\textwidth]{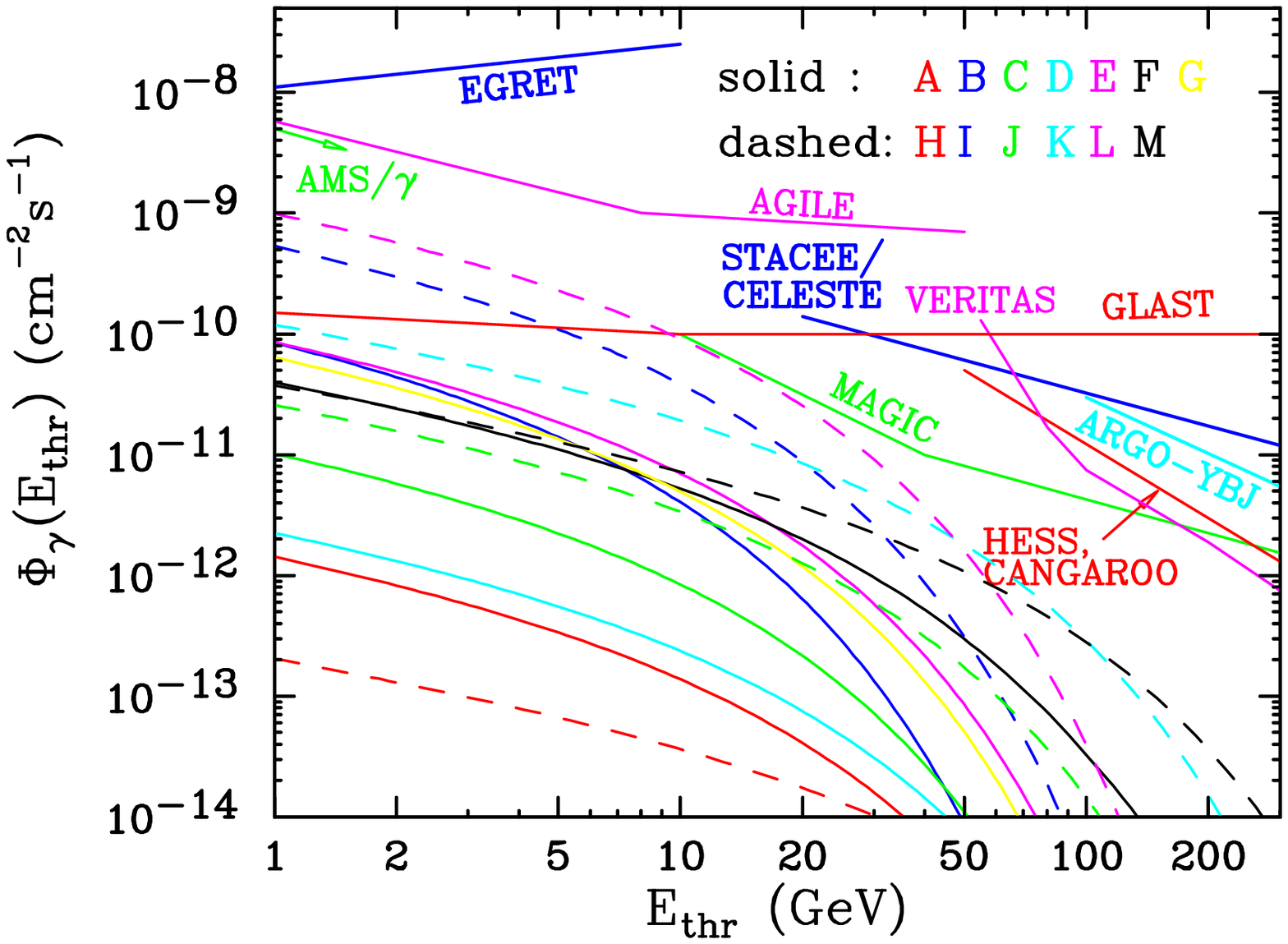}
\end{center}
\caption[]{Fluxes (upper) of muons from the core of the Sun that could be 
detected in a detector deep underground, underwater or in ice, and (lower) of 
photons from the core of the Milky Way, as calculated~\cite{EFFMO} in 
the proposed benchmark scenarios. The calculations are compared with the 
estimated sensitivities of the experiments shown.}
\label{fig:EFFMO}
\end{figure}

A less direct strategy is to look for the products on LSP {\it
annihilations inside the Sun or Earth}. These would produce neutrinos with
relatively high energies, whose interactions in rock would yield muons
that could be detected in a detector deep underground,
underwater~\cite{Thompson} or in ice~\cite{Rhode}. The prospects for
detecting these muons are quite model-dependent, but it seems that
annihilations inside the Sun might be more promising, at least in the
benchmark scenarios~\cite{EFFMO} shown in the upper panel of 
Fig.~\ref{fig:EFFMO}.

Other suggestions have been to look for positrons or antiprotons produced
by LSP {\it annihilations in the halo of the Milky Way}. Quite a large
number of cosmic-ray antiprotons have now been observed, but their flux
and modulation with the solar cycle are consistent with secondary
production by primary matter cosmic rays. As shown here by the AMS
collaboration~\cite{Wallraff}, low-energy positrons in near space are also
mainly produced by the collisions of cosmic rays, in the Earth's
atmosphere, though there are still some hints of a possible anomaly at
higher energies. However, it does not seem possible to reproduce this hint
in the CMSSM~\cite{EFFMO}.

Another suggestion has been to look for $\gamma$-ray production by {\it
annihilations in the core of the Milky Way}, where the relic density may
be enhanced. However, as discussed earlier, there are considerable
uncertainties in this possible enhancement. Depending on its magnitude,
some benchmark scenarios might offer hopes for detection in this
way~\cite{EFFMO,maybe}, as shown in the lower panel of
Fig.~\ref{fig:EFFMO}.

The general conclusion from these benchmark studies in the CMSSM is: think
big! Detectors much larger than the present generation would be required
to have a good chance of detecting elastic scattering, and km$^3$
detectors are probably needed to see annihilations in the Sun or Earth.

\subsubsection{Comments on Neutralino Scattering}

One of the most important contributions to spin-independent elastic
scattering is that due to Higgs exchange, and one of the reasons why
current predictions for the cross section are less optimistic than a few
years ago is the dramatic improvement in the lower limit on the Higgs mass
from LEP. The lower limit of $114.1$~GeV in the Standard Model~\cite{SMH}
also applies to the CMSSM in regions of interest for dark matter, and in
the more general MSSM when $\tan \beta \lappeq 8$~\cite{MSSMH}.

The proton and neutron structure effects on both the spin-dependent and
-independent elastic scattering cross sections are relatively well under
control. Despite residual uncertainties in some relevant hadronic matrix
elements, other uncertainties are probably considerably larger.

Last year, the initial interpretation of the BNL experiment~\cite{BNL} on
$g_\mu - 2$ gave considerable hope to searches for elastic scattering, as
it appeared to exclude large values of $m_{1/2}$ and $m_0$~\cite{oldg-2}.
However, with the recent correction of the sign of the hadronic
light-by-light scattering contribution~\cite{lbl}, the previous
`discrepancy' with the Standard Model prediction for $g_\mu - 2$ has been
greatly reduced, large values of $m_{1/2}$ and $m_0$ are again
allowed~\cite{Nath,Ellis}, and the elastic scattering rate may be very
small, as exemplified by the benchmark studies~\cite{EFFMO}.  However, it
remains true that the rate could be quite large if $g_\mu - 2$ eventually
settles down close to its present central value, and if there are no
further Standard Model surprises in store.

As was discussed here, annual modulation is a potentially powerful tool
for convincing skeptics that a detector signal is indeed due to the
scattering of dark matter, particularly when combined with directional
information~\cite{Vergados}. The DAMA experiment is currently under
pressure from a number of other
experiments~\cite{CDMS,Cline,Juillard,Smith,Giuliani,Seidl,Dawson,Krivosheina}.  
In view of the possible ambiguities in the interpretation of any
experimental signal, it is desirable to explore as many different
techniques as possible, and it was encouraging to hear here that studies
using Sodium Iodide, Germanium, Xenon, Calcium Fluoride, Lithium Fluoride
and Aluminium as target materials are underway. It was also encouraging to
hear that Pulse Shape Discrimination, Time Projection Chambers, Silicon
Drift Detectors, and phonon-based detection strategies are being explored.
It would be particularly impressive to find a confirmatory signal for
spin-dependent scattering, and I recall that Fluorine is the most
promising material for this purpose~\cite{EF}.

\subsubsection{Axino Dark Matter}

\begin{figure}[h]
\begin{center}
\includegraphics[width=.7\textwidth]{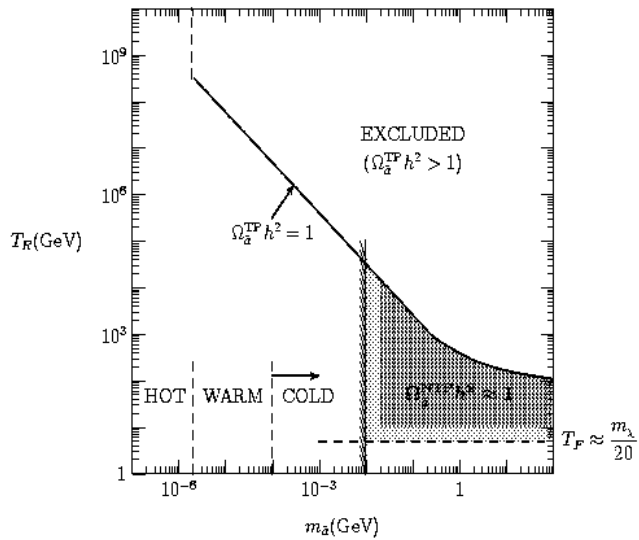}
\end{center}
\caption[]{Compilation of constraints on axino dark matter~\cite{Kim}, 
as functions of the reheating temperature $T_R$ and the axino mass 
$m_{\tilde a}$.} 
\label{fig:axino}
\end{figure}

As discussed here~\cite{Kim}, the axino $\tilde a$ is quite a `naturalÕ
possibility in an extension of the MSSM that includes an axion in order to
explain why the strong interactions conserve CP. The decay constant $F_a$
that reflects the scale of axion dynamics should be $\lappeq 10^{11}$~GeV,
so as to avoid having too much axionic cold dark matter. As in the MSSM, a
plausible mass scale for the lightest neutralino is $m_\chi \sim 100$~GeV.
Assuming these values for $F_a$ and $m_\chi$, Fig.~\ref{fig:axino}
displays the allowed range of the model parameter space in the $m_{\tilde
a}, T_R$ plane, where $T_R$ is the reheating temperature after inflation.
As seen in Fig.~\ref{fig:axino}, one must require $T_R \lappeq 10$~TeV and
$m_{\tilde a} \gappeq 10$~MeV. The latter is not a problem, since models
typically yield $m_{\tilde a} \sim 10$~GeV, but in that case $T_R \lappeq
100$~GeV would be needed, implying a somewhat unconventional cosmology.

\subsubsection{Gravitino Dark Matter?}

The thermal production of gravitinos following inflation has long been
regarded as a potential problem for cosmology. To avoid this, it is
generally considered that the reheating temperature cannot be too high:
$T_R \lappeq 10^9$~GeV. In recent years, the possibility of overproducing
gravitinos during inflation has also been raised~\cite{Linde}. As we heard
here, inflatinos are certainly produced copiously, but these are not
thought to convert into relic gravitinos~\cite{Nilles}.

Since gravitinos have only gravitational-strength interactions, gravitino
dark matter would be unobservable, and hence a nightmare for detection
experiments. In most supergravity scenarios, the gravitino weighs more
than the lightest neutralino, and is unstable. However, the possibility
has been raised of a light gravitino weighing $\sim 1$~KeV, which would be
a potential candidate for warm dark matter. However, this possibility does
not seem to be required, or even favoured, by cosmology, so is not pursued
here.

\subsection{Metastable Superheavy Relics?}

As was discussed by several speakers at this meeting, there may well exist
ultra-high-energy cosmic rays (UHECR) beyond the Greisen-Zatsepin-Kuzmin
(GZK) cutoff, that is expected due to interactions with the
CMB~\cite{Weiler}. The experimental situation is not yet clear, since
there are issues of energy calibration and flux normalization between the
two experiments with the highest statistics, AGASA and Hi-Res. Experiments
at CERN might be able to help by updating studies of fluoresecence due to
particles passing through Nitrogen, and by validating models of
high-energy particle interactions at the LHC.

Assuming that UHECR beyond the GZK cutoff do exist, there are two main
scenarios for their origins: bottom-up mechanisms involving discrete
sources within the GZK range, and top-down mechanisms that invoke the
decays of ultra-massive particles or topological
defects~\cite{Berezinsky}.

Bottom-up mechanisms are clearly more conservative (and hence more
plausible?). They suggest that the observed UHECR should cluster and
perhaps align with sources observable by other means, such as active
galactic nuclei or gamma-ray bursters, unless intergalactic magnetic
fields are strong enough to spread them out. There have indeed been
reports of clustering and possible alignment, but these have yet to attain
consensus.

It has been realized that the gravitational production of superheavy
unstable relics, {\it cryptons}, is likely to be efficient enough to
provide an interesting relic density. Moreover, the hidden sectors of
string models generically contain such particles, which naturally have
long lifetimes, thanks to selection rules and because interactions between
the hidden and observable sectors are usually of very high order and very
weak. As we heard at this meeting~\cite{Fodor}, the crypton decay spectrum 
may fit well
the apparent excess of UHECR beyond the GZK cutoff. However, models tend
to predict a large $\gamma$/proton ratio, whereas the observed UHECR have
been consistent with protons and could not all be photons. In any crypton
model, most of the UHECR would originate from decays within the halo of
the Milky Way. This means that their arrival directions should exhibit a
galactic anisotropy, and could also cluster if much of the halo consists
of clumped cold dark matter.

\begin{figure}[h]
\begin{center}
\includegraphics[width=.5\textwidth]{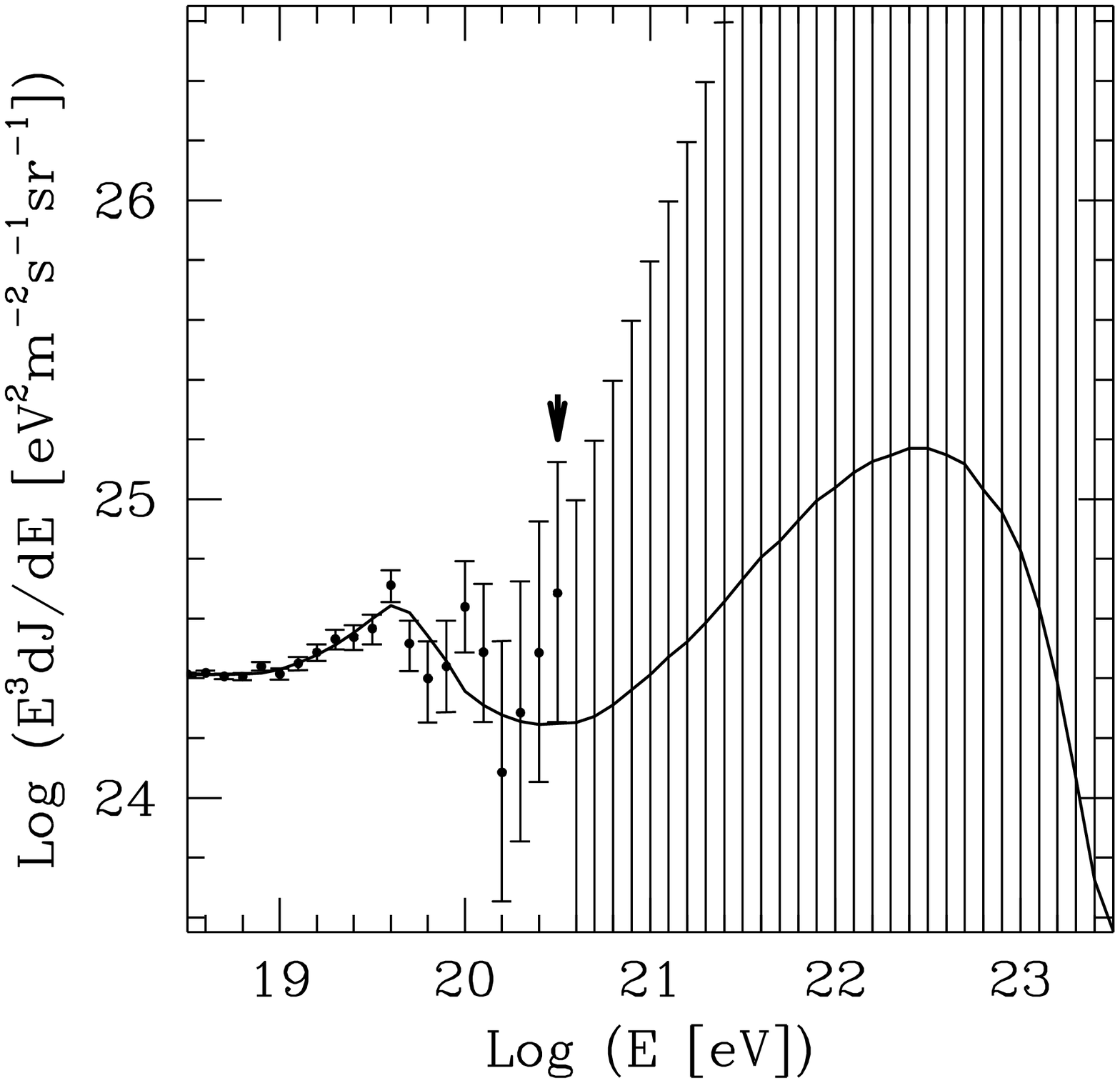}
\includegraphics[width=.5\textwidth]{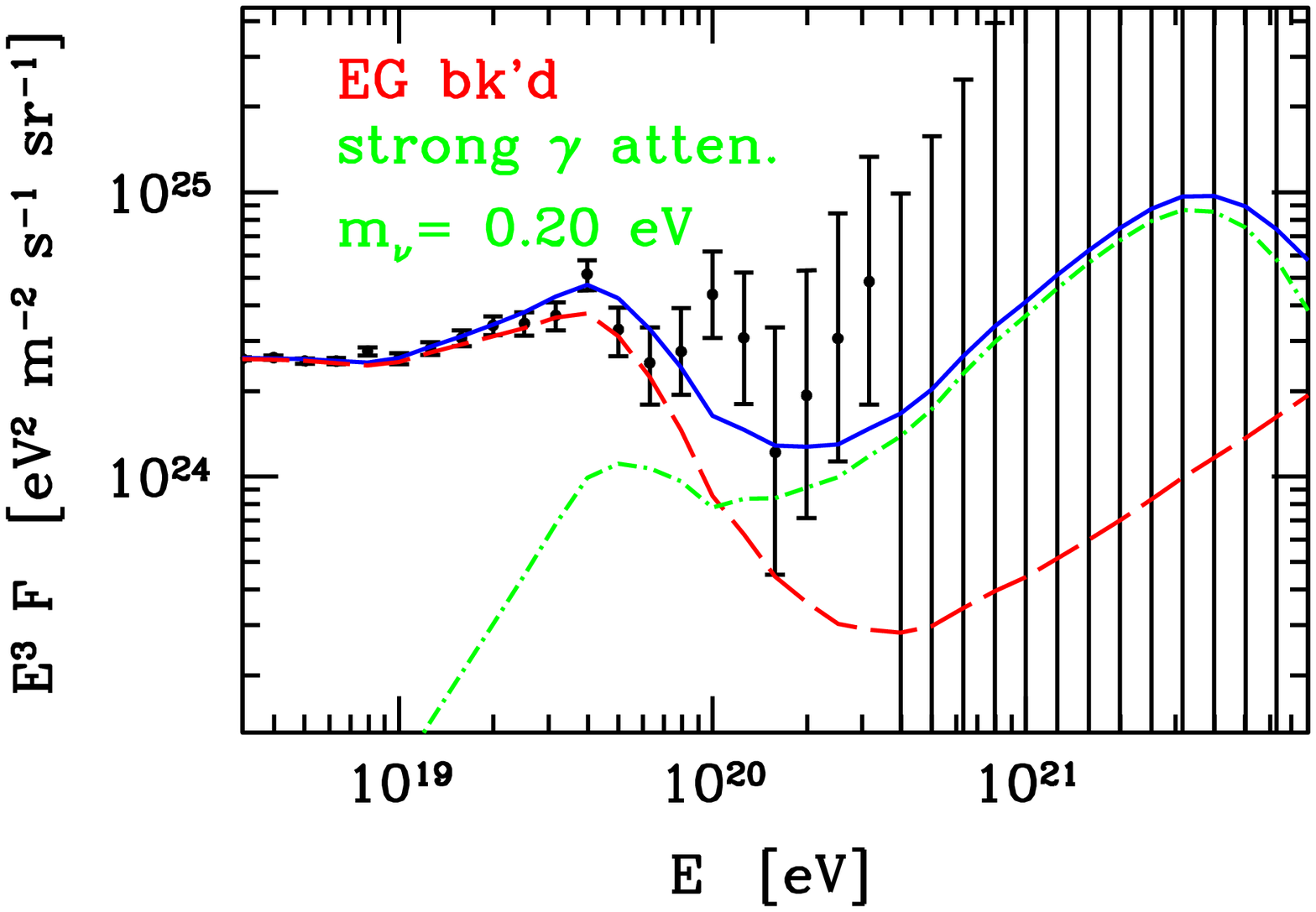}
\end{center}
\caption[]{Illustrations of fits to the UHECR data in the crypton 
decay (upper) and $Z$-burst (lower) models~\cite{Fodor}.}
\label{fig:Fodor1}
\end{figure}

A third scenario discussed at this meeting is the $Z$-burst model,
according to which the UHECR observed originate from collision between
primary UHE neutrinos that strike relic dark matter neutrinos to produce
$Z^0$ bosons that subsequently decay. This scenario requires relic
neutrinos that are somewhat heavier than one might expect on the basis of
atmospheric and solar oscillation experiments, but the hypothesis cannot
be excluded at present, as seen for example in 
Fig.~\ref{fig:Fodor1}~\cite{Fodor}.

As we heard at this meeting, a series of ambitious experiments on UHECR
are now being prepared or planned. The Auger experiment currently being
constructed in Argentina will combine fluorescence and calorimetric
techniques, and have a much larger surface area than previous
experiments~\cite{Auger}. EUSO~\cite{EUSO} and OWL-Airwatch are proposals
for space experiments to look down at fluorescence in the atmosphere over
much larger areas still. The physics issues concerning UHECR are on their
way to being resolved.

\subsection{Alternative Particle Models for Dark Matter}

If all the above, more conventional, particle candidates for dark matter
fall by the wayside, theorists have more exotic candidates in their back
pockets~\cite{Matos,Fayet,Boehm}. For example, there could be interacting
massive fermions that resemble warm dark matter, should that be required.
Alternatively, there could be interacting scalar dark matter. One of these
possibilities might become interesting if the cold dark matter paradigm
fails.

\section{Dark Energy}

As we were reminded repeatedly at this meeting, the cosmological
concordance model seems to require some form of dark energy in `empty'
space, which should make the dominant contribution to the overall energy
density of the Universe:
\begin{equation}
\Omega_\Lambda \; \; = \; \; 0.65 \pm 0.05.
\label{darkenergy}
\end{equation}
Attempts to measure its equation of state suggest that it is almost
constant:
\begin{equation}
\omega_Q \; \; < \; \; 0.7,
\label{quint}
\end{equation}
ruling out some attractive tracking quintessence models~\cite{Bludman}.

We heard at this meeting of some interesting new ideas. For example, the
existence of large extra dimensions would offer new
approaches~\cite{Burgess}, such as obtaining $\Lambda$ from self-tuning in
five dimensions, or radion quintessence in six dimensions. Other ideas
proposed here included a Chaplygin gas~\cite{Tupper} and a repulsive force
in massive QED:  `spintessence'~\cite{Kinney}. Certainly new ideas are
desperately needed.

However, in my view, it makes no sense to discuss dark energy outside the
framework of a complete quantum theory of gravity. Indeed, explaining the
presence and magnitude of dark energy is surely the greatest challenge for
a candidate quantum theory of gravity such as string theory. The Holy
Grail of such a theory should be to calculate the amount of dark energy
from first principles.

\section{Final Remarks}

The quest for dark matter is entering an exciting phase. Astrophysicists
and cosmologists keep on telling us particle physicists that a large
amount of dark matter is certainly required, and that it is probably
mainly cold. There are still some nagging worries about this paradigm,
related to galactic cores and the absence of observable small satellite
galaxies, but these seem resolvable~\cite{Cress} and there is no serious
rival to the cold dark matter paradigm~\cite{Fayet}. However,
astrophysicists cannot tell us what this dark matter is composed of: that
is the task of we particle physicists.

We certainly have plenty of candidates, ranging from axions through
supersymmetric particles to cryptons. Which if any of these is correct can
only be decided by particle experiments, either with accelerators or using
astrophysical sources. The greatest accelerator hopes lie with the LHC,
but non-accelerator experiments have an almost free rein until it starts
taking data in 2007, and some candidates such as axions and cryptons lie
beyond the reach of accelerators.

Dark energy is even more of a challenge than dark matter. Particle
physicists did not expect it, and do not have many convincing ideas for
its origin. A deeper understanding of quantum gravity is surely necessary.

Given this state of ferment, there is plenty to keep us all busy until the
next Dark 200N meeting!

\end{document}